\documentstyle[12pt,aasms4]{article}

\received{}
\accepted{}
\journalid{}{}
\articleid{}{}
\lefthead{Hatano et al.}
\righthead{Global Distribution of Supernovae}
 
\begin{document}
\title{Extinction and the Radial Distribution of Supernova Properties in Their
Parent Galaxies}

\author{KAZUHITO HATANO, DAVID BRANCH, AND JENNIFER DEATON}

\affil{Department of Physics and Astronomy, 
University of Oklahoma, Norman, OK 73019}

\begin{abstract}
                                    
We use a Monte Carlo technique and assumed spatial distributions of
dust and supernova (SN) progenitors in a simple model of a
characteristic SN--producing disk galaxy to explore the effects of
extinction on the radial distributions of SN properties in their
parent galaxies.  The model extinction distributions and projected
radial number distributions are presented for various SN types.  Even
though the model has no core--collapse SNe within three kpc of the
center, a considerable fraction of the core--collapse SNe are
projected into the inner regions of inclined parent galaxies owing to
their small vertical scale height.  The model predicts that because of
extinction, SNe projected into the central regions should on average
appear dimmer and have a much larger magnitude scatter than those in
the outer regions.  In particular, the model predicts a strong deficit
of bright core-collapse events inside a projected radius of a few kpc.
Such a deficit is found to be present in the observations.  It is a
natural consequence of the characteristic spatial distributions of
dust and core--collapse SNe in galaxies, and it leads us to offer an
alternative to the conventional interpretation of the Shaw effect.

\end{abstract}

\keywords{dust, extinction --- galaxies: ISM --- supernovae: general}

\section {Introduction}

Inferring the true radial dependence of supernova (SN) properties in
their parent galaxies from observations of projected radial
distributions is a tricky business.  For example, Bartunov, Makarova,
and Tsvetkov (1992) could find no significant difference between the
radial distributions of SNe of Type Ia (SNe~Ia) and Type II (SNe~II)
in disk galaxies, while van den Bergh (1997) and Wang, H\"oflich, \&
Wheeler (1997) reach opposite conclusions as to whether SNe~Ia or
SNe~II are more centrally concentrated.  Wang et~al also conclude that
SNe~Ia farther than 7.5 kpc from the centers of their parent galaxies
are more homogeneous in absolute magnitude and $B-V$ color than those
that are closer to the centers.  An obvious and generally recognized
difficulty in this kind of work is that only the projected distance of
an SN from the center of its galaxy, not the true distance, is
observed.  Another difficulty, that to our knowledge has not
previously been explored, is that extinction of SNe by dust in their
parent galaxies can affect the observed radial dependence of SN
properties.

To make a first exploration of the effects of extinction on the radial
dependence of SN properties, we have used a Monte Carlo technique
together with assumed spatial distributions of dust and SN progenitors
in a simple model of a characteristic SN--producing disk galaxy.  The
model is described in section 2, and the extinction distributions and
projected radial distributions that it predicts for the various SN
types are presented.  Model predictions of how the brightness
distributions of the various SN types depend on projected radius are
presented and compared with observation in section 3.  Section~4
presents a brief discussion.

\section {The Model} 

\subsection {Input}

Our Monte Carlo technique for studying the visibility of SNe and novae
(Hatano et~al. 1997a,b,c) was inspired by a study of the visibility of
Galactic SNe by Dawson and Johnson (1994).  In their model and in
Hatano et~al. (1997a,b), which were concerned with SNe and novae in
the Galaxy, a double exponential distribution of dust was used, with a
radial scale length of five kpc and a vertical scale height of 0.1
kpc.  Core--collapse events (SNe~II and Ibc) were distributed like the
dust except that they were not allowed to occur within three kpc of
the center in view of the lack of evidence for much recent star
formation in the central regions of our Galaxy.  SNe~Ia consisted of
two spatially interpenetrating components: disk SNe~Ia had a double
exponential distribution with a scale length of five kpc and a
vertical scale height of 0.35 kpc and they were not truncated at three
kpc.  Bulge SNe~Ia were spherically distributed (in kpc) as $(R^3 +
0.7^3)^{-1}$.  Disk SNe~Ia outnumbered bulge SNe~Ia by a factor of
seven, consistent with the estimated Galactic disk--to--bulge mass
ratio (van der Kruit 1990).  The bulge was truncated at three kpc and
the disk at 20 kpc.  For more detailed descriptions of the model, see
Dawson and Johnson (1994) and Hatano et al. (1997a).

Hatano et al. (1997c) was concerned with novae in M31.  Because the
dust density in M31 is known to peak not at the center, but well out
in the disk where most of the current star formation is taking place,
the model dust distribution was modified accordingly.  Recently
Sodroski et al. (1997) have found that in the Galaxy, too, the dust
density peaks off center, at a radius of about 5 kpc, where the total
face--on extinction through the disk is $A_B \simeq 1$.  For the
present study we have made a simple parameterization of the radial
dependence of the total face--on extinction through the disk, after
Figure 9 of Sodrowski et~al.  This radial dependence of the dust,
together with a retained vertical scale height of 0.1 kpc, leads to
the following expression for the $B$--band extinction per kpc,
$\alpha$, in the plane of the disk:

$$\alpha(r) = r,\ \ \ \ r \le 5\ \rm kpc,   \eqno (1) $$
$$\alpha(r) = -0.4r + 7,\ \ \ \ r > 5\ \rm kpc   \eqno (2) $$   

\noindent Thus the dust density rises from zero at the center of the
model to a maximum value at 5 kpc and then falls to zero at 17.5 kpc.
This particular parameterization gives a rather high value of 3.8 mag
kpc$^{-1}$ in the plane of the disk at $r=8$ kpc, i.e., at the radial
distance of the sun in the Galaxy, but considering that large galaxies
produce more SNe and tend to be more dusty than small ones (van den
Bergh \& Pierce 1990), it probably is not unreasonable for this
initial exploration of the effects of extinction.

\subsection {Model Extinction Distributions}

Figure~1 shows the model extinction distributions for bulge SNe~Ia,
disk SNe~Ia, and core--collapse SNe, each for five different galaxy
inclinations as well as for a random distribution of inclinations.
Note that for any inclination, even edge--on, the distributions for
bulge and disk SNe~Ia are strongly peaked at small values of $A_B \le
0.1$. (For the adopted disk--to--bulge SN~Ia ratio of seven, the total
distribution for SNe~Ia is much like the distribution for disk
SNe~Ia.) Core--collapse SNe tend to be more extinguished than SNe~Ia
owing to their smaller vertical scale height.

Table 1 quantifies the model extinction distributions.  Column (1)
gives the cosine of the inclination of the model, where $\cos(i)=1$
refers to the face--on case.  Columns (2)--(4) refer to bulge SNe~Ia
(Ia--b) and list the mean extinction $<A_B>$, the mean extinction of
an ``extinction--limited subset'' of bulge SNe~Ia that have $A_B <
0.6$, and the fraction of the events in that subset.  Columns (5)--(7)
give the same information for disk SNe~Ia (Ia--d) and columns
(8)--(10) are for core--collapse SNe (CC).  In the edge--on case the
mean extinction of all SN types becomes high, but of course there
would be a strong observational selection acting against the discovery
of severely extinguished events.  It is interesting that the mean
extinction of the extinction--limited subsets {\sl decreases} as
inclination increases, i.e., when the model is highly inclined the
extinction tends to be all or nothing.  Note that even in the edge--on
case substantial fractions of bulge and disk SNe~Ia make it into the
extinction--limited subsets, while only a small fraction of the
core--collapse SNe do.  Column (10) shows that, to the extent that an
extinction of 0.6 mag is sufficient to significantly reduce the
probability that a SN is discovered, this particular model is dusty
enough to support an inclination--dependent observational
discrimination against the discovery of core--collapse SNe (cf. van
den Bergh 1993; Tammann 1994; Cappellaro \& Turatto 1997).

\subsection {Model Projected Radial Distributions} 

Our technique allows us to plot model projected radial distributions
of SNe for comparison with observed distributions.  The model
distributions presented here have been calculated with a disk
truncation radius of 30 rather than 20 kpc because some SNe have been
observed beyond 20 kpc.

Figure~2 shows normalized projected radial distributions for disk
SNe~Ia and core--collapse SNe, for five inclinations and a random
distribution of inclinations. As the inclination increases, the
projected distributions shift toward the central regions. The behavior
of the core--collapse SNe is especially interesting: in the face--on
case, by construction, there are no core--collapse events within three
kpc, but in the edge--on case plenty are projected into the innermost
regions because of the small vertical scale height.  This illustrates
that the observational presence of core--collapse events projected
near the center does not guarantee the actual existence of
core--collapse events near the center.  More generally, Figure~2
illustrates the difficulty of inferring details about the relative
radial distributions of SN types from observations of their projected
radial distributions in samples that include highly inclined galaxies,
especially if the SN types have different vertical scale heights.

van den Bergh (1997) studied projected SN radial distributions in a
sample of disk galaxies restricted to those having $i < 70^o$, to
minimize extinction and projection effects, and he tentatively
concluded that SNe~Ia are more centrally concentrated than SNe~II.
Wang et~al (1997) presented calculated projected radial distributions,
analagous to our Figure~3 but with very different model parameters.
They compared to observed distributions and tentatively concluded the
opposite --- that SNe~II are more centrally condensed than SNe~Ia.
Because Wang et~al. did not use a smaller vertical scale height for
SNe~II than for SNe~Ia, they probably overestimated the true central
concentration of SNe~II relative to that of SNe~Ia; thus it appears to
us that van den Bergh's conclusion is more likely to be correct.

\section {The Radial Dependence of SN Brightness}

At this point we have to specify the SN luminosity functions.  After
Hatano et~al (1997a) we adopt gaussian distributions of the intrinsic
absolute magnitudes of SNe~Ia, Ibc, and II, with mean absolute
magnitudes of -19.5, -18.0, and -17.0, and dispersions of 0.2, 0.3, and 1.2, respectively.

\subsection {Type Ia Supernovae}

The top panel of Figure~3 shows the model distribution of absolute
magnitude as viewed by the external observer who has not corrected for
parent--galaxy extinction (i.e., the quantity $M_B + A_B$) for SNe~Ia.
Those at large projected radial distances are practically
unextinguished, while some of those at smaller projected distances are
severely extinguished. The lower panel of Figure~3 shows a
corresponding observational plot, constructed from data in the June 6,
1997 version of the Asiago Supernova Catalogue with distances based on
parent--galaxy radial velocity and $H_0 = 60$ km s$^{-1}$ Mpc$^{-1}$.
Apparent magnitudes have been corrected for foreground extinction in
the Galaxy but not for extinction in the parent galaxies.  Considering
the simplicity of our model, and that the lower panel is affected by
strong observational bias against the discovery of severely
extinguished events, the general resemblance of the two panels of
Figure~3 is satisfactory.  Figure~3 strongly suggests that the effects
of parent--galaxy extinction are likely to be the cause of much of the
observed difference in absolute--magnitude dispersion between SNe~Ia
within and beyond 7.5 kpc, a difference first pointed out by Wang
et~al. (1997).  It should be noted, however, that for a small sample
of SNe~Ia having high--quality data, Wang et~al. also found an excess
of {\sl overluminous} SNe~Ia within 7.5 kpc, which of course cannot be
explained in terms of extinction.  Wang et~al also called attention to
an apparent observational deficit of SNe~Ia in the innermost one kpc
of their parent galaxies.  The upper panel of Figure~3 shows that our
model predicts the presence of observationally bright SNe~Ia (mostly
bulge SNe~Ia) projected into the central kpc, while none appear in the
observational plot (lower panel of Figure~3).  This supports the
suggestion by Wang et~al. that bulge populations may be poor producers
of SNe~Ia.  (The conclusion of Wang et~al that SNe~Ia therefore do not
come from novae does not follow for us, because we [Hatano
et~al. 1997b,c] have argued that Galactic and even M31 novae do not
come primarily from the bulge populations as is usually assumed.)

\subsection {Core--Collapse Supernovae}

Figure 4 is like Figure~3, but for SNe~II.  Although the effects of
the broad adopted luminosity function for SNe~II are apparent, the
upper panel of Figure 4 shows the same general tendency as Figure~3
for many severely extinguished model SNe to appear at relatively small
projected distances from the center.  (Some are extinguished right off
the top of Figure~4.) The large triangle in the upper panel serves to
call attention to predicted deficit of bright SNe~II in the central
regions.  In the model the deficit reflects the tendency of SNe~II
that are projected into the central regions to be substantially
extinguished.  Although more data are needed to reach a final
conclusion, this deficit does appear to be present in the observations
(lower panel). In the same way the model predicts a deficit of bright
SNe~Ibc in the central regions, and the inset to the lower panel shows
that although the numbers are still smaller, observations of SNe~Ibc
are at least consistent with the presence of the deficit.

We have explored the consequences for the core--collapse deficit of
altering the dustiness of the model.  Halving the dustiness makes the
model deficit less pronounced than the observational one, while
doubling the dustiness extinguishes nearly all centrally projected
core--collapse SNe beyond detectability.  Measuring the radial
dependence of controlled samples of SNe in galaxies appears to be a
promising way to constrain the amount and the distribution of dust in
galaxies, which is a controversial issue (Valentijn 1990; Burstein,
Haynes, \& Faber 1991).

The results shown in Figure 4 lead us to suggest an alternative to the
classical explanation of the selection effect discussed by Shaw (1979)
--- the observational deficit of SNe in the central regions of remote
galaxies relative to nearer galaxies.  (See van den Bergh (1997) and
Wang et~al (1997) for plots that illustrate the Shaw effect with
present data.)  Shaw's interpretation, which has been generally
accepted, was that the effect is caused by the increasing difficulty,
with distance, of discovering SNe against the bright central regions
of galaxies.  But if this is the primary cause of the Shaw effect then
in Figure~4 we would expect to see a deficit of observationally {\sl
dim} SNe in the central regions, because they would be more readily
lost than bright ones. Then the large triangle in the lower panel of
Figure~4 should be upside down.  Instead, the lower panel of Figure 4
shows a deficit of observationally {\sl bright} SNe~II in the central
regions, as predicted by our model.  Therefore we suggest that the
Shaw effect for SNe~II (and SNe~Ibc) comes about at least in part
because core--collapse SNe projected into the central regions of
galaxies tend to be observationally dim, and the difficulty of
discovering dim SNe increases with distance.  Our model does not
predict such a deficit for SNe~Ia, but it is not so clear that there
really is an observational Shaw effect for SNe~Ia (see Figure~3 of van
den Bergh [1997] and Figure~1 of Wang et~al [1997]).

\section {Discussion}     

Our simple model of the spatial distributions of dust and SN
progenitors in a characteristic SN--producing disk galaxy predicts
that owing to extinction, SNe projected into the inner regions of
their parent galaxies tend to be observationally dimmer and to have a
much larger magnitude dispersion than SNe in the outer regions.  As
Wang et~al (1997) first showed, such is the case for SNe~Ia.  In our
model the effect is severe enough, for core--collapse SNe, to produce
a deficit of observationally bright events projected into the central
regions.  For core--collapse SNe, such a deficit does appear to be
present in the observational data.  Although the need for improving on
the assumption of a single characteristic SN--producing disk galaxy in
future work is obvious, this effect of extinction is likely to be
real.  It has obvious implications for such things as using
high--redshift SNe~Ia to determine the cosmic deceleration (avoid
extinction by using events in the outer regions of their parent
galaxies); for determining the true relative radial distributions of
SN types (extinction can cause confusion, especially when comparing SN
types that have different vertical scale heights), and for determining
SN rates (substantial corrections for incompleteness that depend on
parent--galaxy dustiness are required).

\acknowledgements 

Adam Fisher wrote the original version of our Monte Carlo code.  
We are grateful to Eddie Baron, Darrin Casebeer, Dean Richardson, Bill
Romanishin, and Rollin Thomas for many discussions in the course of
this work, which was supported by NSF grant AST 9417102.  J.D. was
supported in part by an NSF REU Supplement to AST 9417242.

\clearpage

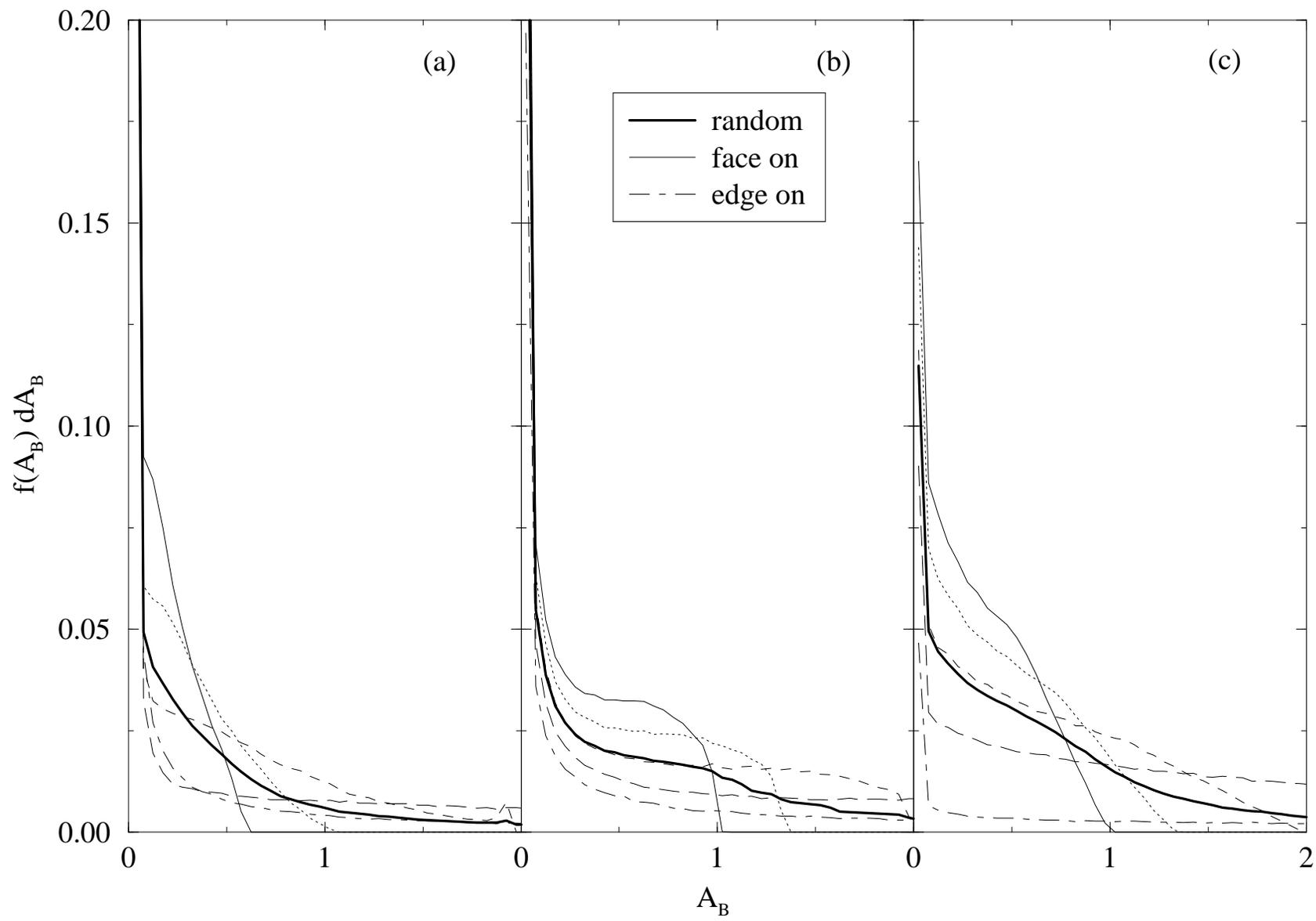
\begin{figure}
\figcaption{The model $A_B$ distributions for (a) bulge SNe~Ia; (b)
disk SNe~Ia; and (c) core--collapse SNe, each for five inclinations of
the model.  The inclinations are equally spaced in cos(i).  The thick
solid lines are the $A_B$ distributions for a random distribution of
inclinations.
\label{fig1}} \end{figure}

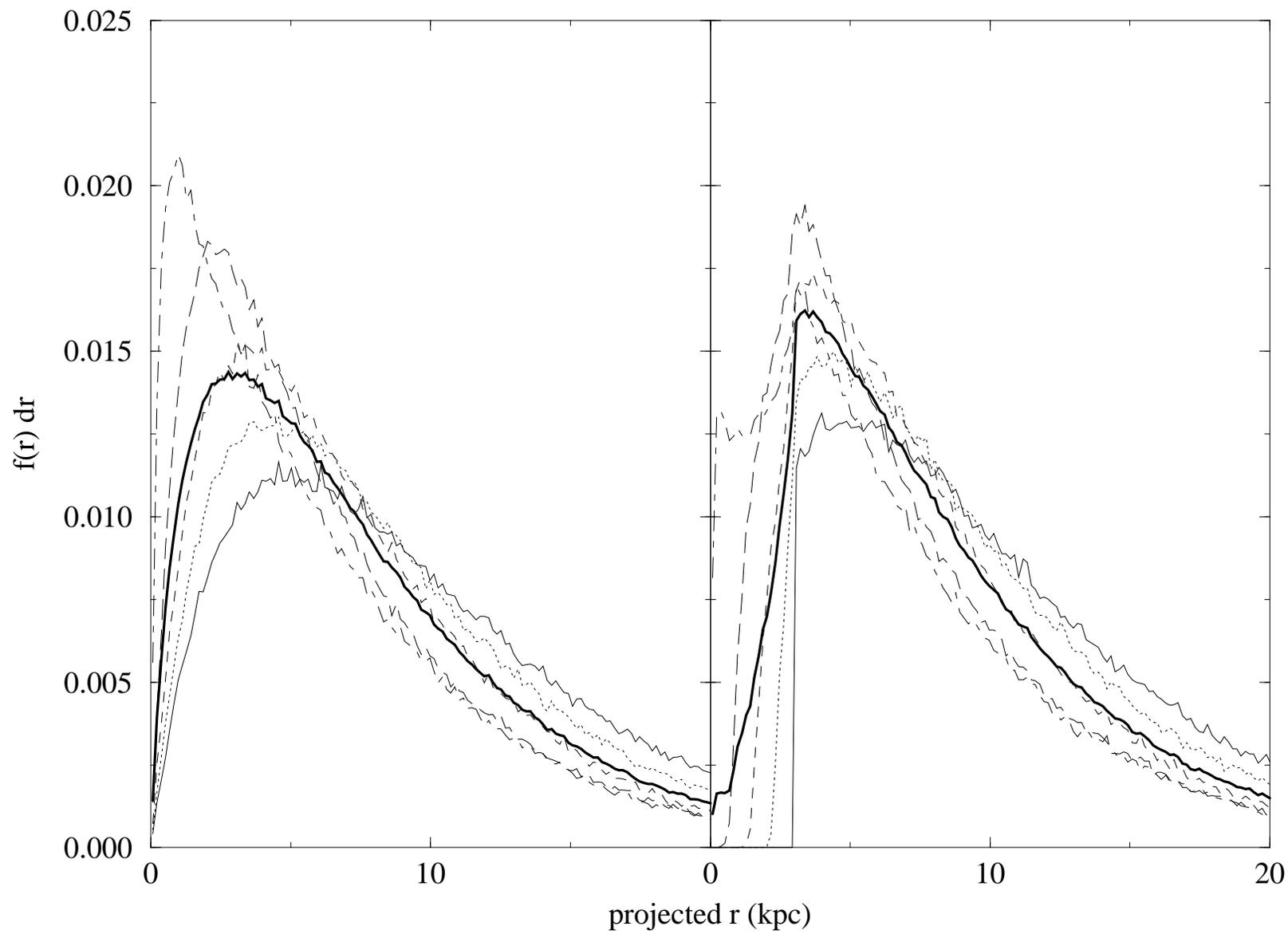
\begin{figure}
\figcaption{Normalized model projected radial distributions of disk SNe~Ia and
core--collapse SNe, for five inclinations and for random inclinations.
Line legends are as Figure 1.
\label{fig2}} \end{figure}

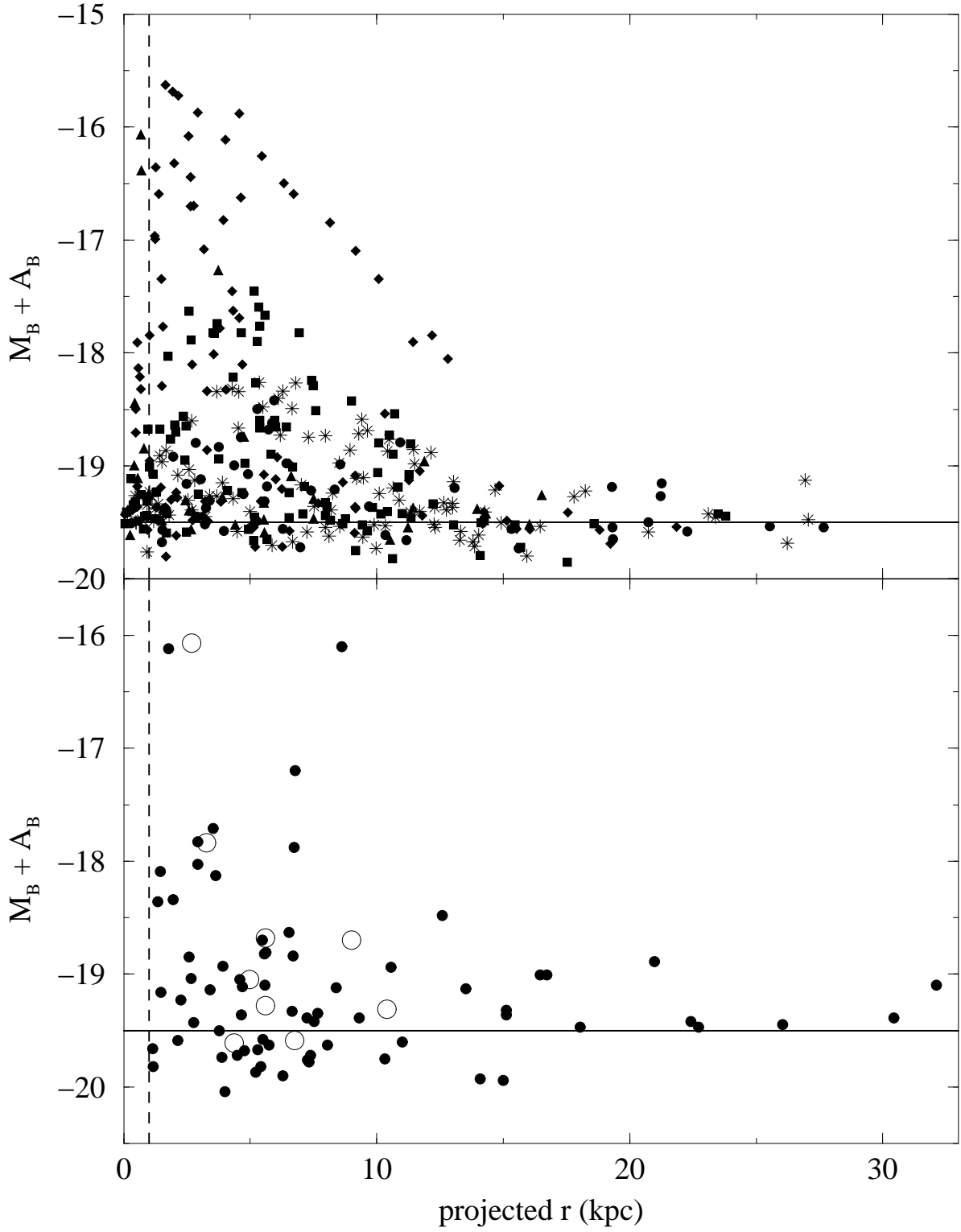
\begin{figure}
\figcaption{{\sl upper}: The model distribution of $M_B + A_B$ for SNe~Ia
plotted against the projected radial distance from the center of the
model.  The distribution of inclinations is random, with symbols as
follows: cos(i) = 1 -- filled circle; cos(i) = 0.75 -- asterisk;
cos(i) = 0.5 -- square; cos(i) = 0.25 -- diamond; and cos(i) = 0,
triangle.  The horizontal line corresponds to the mean of the adopted
intrinsic absolute--magnitude distribution for SNe~Ia and the dashed
vertical line as at one kpc {\sl lower}: The corresponding observed
distribution based on data from the Asiago Supernova Catalogue.  Open
circles denote peculiar SNe~Ia, two of which are intrinsically red.
\label{fig3}} \end{figure} 

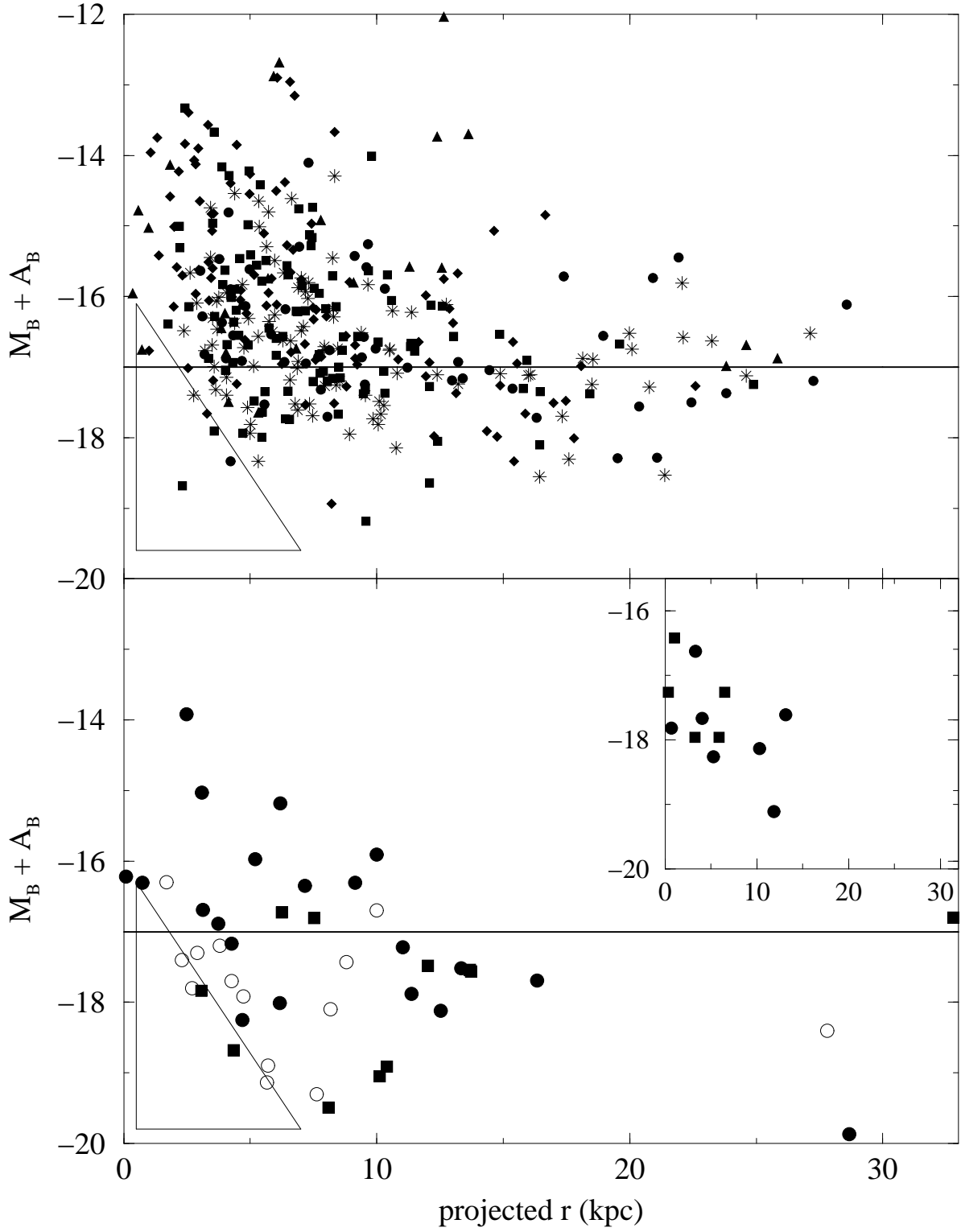
\begin{figure} 
\figcaption{Like Figure 3, but for SNe~II.  In the upper panel, symbols
are as in Figure~3. In the lower panel, squares denote SNe~II--P,
filled circles denote SNe~II--L, open circles denote peculiar SNe~II.
The large triangles call attention to the predicted and observed
deficits of observationally bright SNe~II in the central regions.  The
inset in the bottom panel is for observed SNe~Ibc.  Circles denote SNe~Ib and 
squares denote SNe~Ic.  
\label{fig4}} \end{figure} 

\clearpage

\begin{deluxetable}{cccccccccc}
\footnotesize
\tablenum{1}
\tablecaption{Extinction of Various Supernova Types}
\tablewidth{0pt}
\tablehead{
\colhead{(1)} & \colhead{(2)} & \colhead{(3)} & \colhead{(4)} & \colhead{(5)} & \colhead{(6)} & \colhead{(7)} & \colhead{(8)} & \colhead{(9)} & \colhead{(10)} 
\\
\colhead{} & \colhead{Ia--b} & \colhead{Ia--b} & \colhead{Ia--b} & \colhead{Ia--d} & \colhead{Ia--d} & \colhead{Ia--d} & \colhead{CC} & \colhead{CC} & \colhead{CC} 
\\
\colhead{cos(i)} & \colhead{$<A_B>$} & \colhead{$<A_B>$} & \colhead{f(Ia--b)} & \colhead{$<A_B>$} & \colhead{$<A_B>$} & \colhead{f(Ia--d)} & \colhead{$<A_B>$} & \colhead{$<A_B>$} & \colhead{f(CC)}       
}

\startdata
1.00 & 0.12 & 0.12 & 1.00 & 0.30 & 0.16 & 0.76 & 0.32 & 0.24 & 0.84 \\
0.75 & 0.18 & 0.14 & 0.92 & 0.40 & 0.15 & 0.65 & 0.41 & 0.24 & 0.71 \\
0.50 & 0.36 & 0.13 & 0.72 & 0.58 & 0.14 & 0.55 & 0.62 & 0.24 & 0.54 \\
0.25 & 1.06 & 0.08 & 0.46 & 1.17 & 0.13 & 0.43 & 1.21 & 0.22 & 0.34 \\
0.00 & 4.26 & 0.07 & 0.52 & 8.80 & 0.12 & 0.32 & 19.8 & 0.14 & 0.07 \\
\enddata
\end{deluxetable}


\clearpage

\begin{references} 

\reference{} Bartunov, O. S., Makarova, I. N., and Tsvetkov, D. Yu. 1992, A\&A, 264, 428

\reference{} Burstein, D., Haynes, M. P., and Faber, S. M. 1991,
Nature, 353, 515

\reference{} Cappellaro, E. \& Turatto, M. 1997, in Thermonuclear
Supernovae, ed. R. Ruiz--Lapuente, R. Canal, \& J. Isern (Dordrecht:
Kluwer), 77

\reference{} Dawson, P. C. \& Johnson, R. G. 1994, J. R. Astron. Soc. Can., 88, 369

\reference{} Hatano, K., Fisher, A., and Branch, D. 1997a, MNRAS, 290, 360

\reference{} Hatano, K., Branch, D., Fisher, A., and Starrfield,
S. 1997b, MNRAS, 290, 113

\reference{} Hatano, K., Branch, D., Fisher, A., and Starrfield,
S. 1997c, ApJL, 487, L45

\reference{} Shaw, R. L. 1979, A\&A, 76, 188

\reference{} Sodroski, T. J., Odegard, N., Arendt, R. G., Dwek, E.,
Weiland, J. L., Hauser, M. G., \& Kelsall, T. 1997, ApJ, 480, 173

\reference{} Tammann, G. A. 1994, in Supernovae, ed. S. A. Bludman,
R. Mochkovitch, \& J. Zinn--Justin (Amsterdam: North Holland), 1

\reference{} Valentijn, E. A. 1990,  Nature, 346, 153

\reference{} van den Bergh, S. 1993, Comments Ap., 17, 125 

\reference{} van den Bergh, S. 1997, AJ, 113, 197

\reference{} van den Bergh, S. \& Pierce, M. J. 1990, ApJ, 364, 444. 

\reference{} van der Kruit, P. 1990, in The Milky Way as a Galaxy,
ed. R. Buser \& I. King (Mill Valley: University Science Books), 331

\reference{} Wang, L., H\"oflich, P., and Wheeler, J. C. 1997, ApJ, 483, L29

\end{references}
\end{document}